\documentclass{article}
\pdfoutput=1

\newif\ifsok\soktrue

\usepackage{a4wide}

\usepackage{color}\definecolor{light-gray}{gray}{0.95}
\usepackage[table]{xcolor}

\usepackage{float}
\floatstyle{ruled}\restylefloat{figure}  

\usepackage{caption}
\usepackage{amsmath}
\usepackage{amsfonts}

\usepackage{verbatim}
\usepackage{authblk}
\usepackage{url}
\usepackage{wrapfig}
\usepackage{subcaption}

\usepackage{multirow}

\usepackage{tikz}\usetikzlibrary{shapes,snakes,shadows,calc,automata,positioning}

\newcommand{\EEncSymb}{\mathsf{Enc}}
\newcommand{\pk}{\mathit{pk}}
\newcommand{\EEnc}[2][\pk]{\EEncSymb(#1,#2)}

\newcommand{\GZero}{\mathsf{G}}
\newcommand{\GZeroPrime}{\mathsf{G}'}
\newcommand{\GOne}{\mathsf{H}}
\newcommand{\GTwo}{\mathsf{BS}}
\newcommand{\GThree}{\mathsf{IV}}
\newcommand{\GFour}{\mathsf{UV}}

\title{A short introduction to secrecy \\ and verifiability for elections}

\author[1]{Elizabeth A. Quaglia}

\affil[1]{Information Security Group, Royal Holloway,\authorcr University of London, UK}

\author[2]{Ben Smyth}
\affil[2]{Interdisciplinary Centre for Security, Reliability and\authorcr Trust, University of Luxembourg, Luxembourg}

\date{September 19, 2018}

\newcommand{\heliosspec}{Helios'12}

\begin{document}

\maketitle

\setlength{\textfloatsep}{1em}

\emph{
Many election schemes rely on art, 
rather than science, to ensure that choices are made freely and 
with equal influence. Such schemes build upon creativity and skill,
rather than %theoretical, 
scientific foundations. These schemes are typically 
broken in ways that compromise free choice, 
e.g.,~\cite{Rop07:NetherlandsVoting,DebraBowenCalifornia07,Halderman10:IndiaVoting,Halderman12:DCVoting,Halderman14:EstoniaVoting},
or permit adversaries to unduly influence the outcome, 
e.g.,~\cite{JonesSimons12:VotingBook,GermanyCourt09,DebraBowenCalifornia07,ElectoralCommision07}.
This article shows how such breaks can be avoided by carefully formulating security 
definitions, and proving that schemes satisfy these definitions. 
Equipped with these definitions, we can build election schemes that can be proven to behave as expected.} 

\begin{figure}[t]\caption{Rediscovering verifiability: A historic perspective on election scheme evolution}

Making choices in private has not always been the way. 
``Americans used to vote with their voices -- \emph{viva voce} -- or with their hands or with their feet. Yea or nay. Raise your hand. All in favor of Jones, stand on this side of the town common; if you support Smith, line up over there"~\cite{Lepore08}.
Voting in public naturally enables election verifiability, since each voter can compute
the outcome themselves. But,
\ifsok Mill~\cite{Mill1830} eloquently argues that \fi 
 choices cannot be expressed freely in public: 
``The unfortunate voter is in the power of some opulent man; the opulent man 
informs him how he must vote. Conscience, virtue, moral obligation, religion, 
all cry to him, that he ought to consult his own judgement, and faithfully 
follow its dictates. The consequences of pleasing, or offending the opulent
man, stare him in the face; ... %the oath is violated, 
the moral obligation is disregarded, a faithless, a prostitute, a pernicious 
vote is given\ifsok."\else"~\cite{Mill1830}.\fi\ 
To ensure social constraints are kept at bay, voting became a private act.
In particular, a voter typically marks their choice on a ballot paper
in the isolation of a polling booth and deposits their marked paper into 
a locked ballot box. The isolation of the polling booth is intended to 
facilitate free choice at the time of marking. Moreover, privacy is preserved 
during tallying by mixing the ballot papers prior to counting.
And ``this idea has become the current \emph{doxa} of democracy-builders worldwide"~\cite{Bertrand07:Opening:HiddenHistorySecrectBallot}.
But, 
unlike raising hands, voters cannot be assured that ballots 
are counted correctly.
Nonetheless, the transparency of the whole election process from ballot casting to tallying
and the impossibility of altering the markings on a paper ballot sealed inside a locked 
ballot box gives an assurance of correctness. 
Transparency is lost in electronic election schemes, because software and hardware are used to 
construct ballots and transmit them over public communication channels, and it is difficult 
to observe electronic operations performed on bitstrings. Consequently, choices might be 
altered in ways that cannot be detected.
This led to the rediscovery of verifiability, which has become an essential 
requirement~\cite{VoteFoundation15}.
\end{figure}

An election scheme is a decision-making mechanism to choose a representative\ifsok~\cite{Lijphart84,Saalfeld95,Gumbel05:StealThisVote,Alvarez10:ElectronicElections}\fi,
typically consisting of at least the following three steps.
First, an administrator initialises the scheme (\emph{setup}). 
Secondly, each voter constructs and casts a ballot for their choice (\emph{voting}). 
These ballots are authenticated and recorded using a mechanism, e.g., a bulletin board.
Thirdly, the administrator 
tallies the recorded ballots and
announces an outcome, i.e., a frequency distribution of choices (\emph{tallying}).
This distribution is used to select a representative. For example, in first-past-the-post election schemes
the representative corresponds to the choice with highest frequency.

Choices must be made freely, 
\ifsok
which can be achieved by making choices in private~\cite{UN:HumanRights,OSCE:HumanRights,OAS:HumanRights}%
\else 
which the United Nations%
%UN
\ifsok~\cite[Article 21]{UN:HumanRights}\fi,
the Organization for Security \& Co‑operation in Europe%
%OSCE
\ifsok~\cite[Paragraph 7.3]{OSCE:HumanRights}\fi\ and 
the Organization of American States%
%OAS
\ifsok~\cite[Article 23]{OAS:HumanRights}\fi\
 recommend achieving by making choices in private%
\fi, i.e., 
``when numerous social constraints in which citizens are routinely and
  universally enmeshed -- community of religious allegiances, the patronage of big
  men, employers or notables, parties, `political machines' -- are kept at bay"~\cite{Bertrand07:Opening:HiddenHistorySecrectBallot}. 
This has led to the emergence of the following requirement.

\begin{itemize}
\item Ballot secrecy: a voter's choice is not revealed to anyone.
\end{itemize}

\noindent
Ballot secrecy ensures that a voter's choice is kept secret, which is intended to prevent unwanted consequences (including the preclusion of free choice) that might otherwise arise. 

To illustrate how ballot secrecy can be achieved, we introduce a simple election scheme
that instructs voters to encrypt their choices and instructs administrators to 
decrypt encrypted choices to obtain the outcome. More specifically, 
the scheme works as follows: first, the administrator generates a public key. Secondly, each voter encrypts their choice using that key. Finally, the administrator decrypts each encrypted choice and outputs the corresponding outcome.
Intuitively, ballot secrecy is achieved if the underlying encryption scheme is secure, 
i.e., the encryption of a choice leaks no information about that choice.

Voters, and any other interested parties, must be able to convince
themselves that the announced outcome
is indeed the distribution of choices made by voters,
which can be achieved by making elections verifiable, i.e., ensuring 
``there [is] enough evidence for anyone 
who doubts the results to re-examine and rationally determine whether the 
[outcome was] called correctly"~\cite{VoteFoundation15}.
Election verifiability can be captured by the following  
requirements.

\begin{itemize}
\item Individual verifiability:
	voters can check that the ballots they constructed are recorded. 

\item Universal verifiability:
	anyone can check that the announced outcome is the distribution of voters' choices expressed in the recorded ballots.
\end{itemize}

\noindent 
Taken together, these properties intuitively ensure that anyone can convince themselves that 
the announced outcome corresponds to the choices expressed in the recorded ballots,
and voters can convince themselves that their ballot is included amongst the ballots recorded, hence, their choice is included in the outcome announced by the administrator.\ifsok%
\fi\ 
Election verifiability requires election schemes to provide an additional (\emph{verification}) step to perform the necessary checks.

Verifiability is not ensured by our election scheme based 
upon encryption. Indeed, a spuriously announced outcome need not 
even correspond to the encrypted choices! We introduce a simple election scheme
that achieves verifiability. The scheme instructs each voter to pair their choice with a random value
(i.e., a nonce) and instructs the administrator to compute the election outcome
from those pairs. This scheme ensures verifiability, because
voters can use their nonce to check that their ballot is recorded
(individual verifiability) and anyone can recompute the election outcome
to check that it corresponds to votes expressed in recorded ballots
(universal verifiability). But, ballot secrecy is not ensured, because all 
votes are revealed. 
To simultaneously satisfy both
secrecy and verifiability, more advanced schemes are required.

A rich selection of election schemes have been proposed in the research literature. 
One of the most prominent schemes is
\emph{Helios}~\cite{AdidaPereiraMarneffeQuisquater}, an open-source, web-based election system. 
The notoriety of Helios is partly due to its elegant construction and use in binding elections.
For instance, by the ACM,  the International Association of Cryptologic Research, the Catholic University of Louvain, and Princeton University. 
The scheme works as follows:
first, an administrator generates a public key and a proof of correct key construction (setup).
Secondly, each voter encrypts their choice with the public key, proves correct ciphertext construction, and casts the ciphertext coupled with the proof as their ballot (voting).  
Thirdly, the administrator collects the ballots cast, discards any ballot for which proofs do not hold, homomorphically combines the ciphertexts in the remaining ballots
to derive the encrypted outcome,
decrypts, proves correctness of decryption, and announces the outcome and proof (tallying).
Finally, any interested party recomputes the aforementioned combination and verifies all proofs, and voters verify that
the ballots they constructed are amongst those collected (verification).
Helios was first implemented as Helios 2.0. It is intended to satisfy
ballot secrecy due to encryption, and 
election verifiability because encryption and decryption steps are accompanied 
by proofs. 

%\floatstyle{ruled}
%\restylefloat{figure}
%\begin{figure}\caption{Let's play a game: Formulating security properties}
One way to evaluate security of a scheme is to formulate the desired security properties and check whether the scheme satisfies them.
Cryptographers formulate security properties using \emph{games}~\cite{Katz07}.
Typically, a game consists of a series of interactions between a benign challenger and a malicious adversary. 
An adversary wins a game if it successfully completes a challenge set by the challenger (e.g., distinguish between two scenarios).
Winning captures an execution of the scheme in which the desired security property does not hold. Thus, when formulating a game, the challenge captures what the adversary should not be able to achieve.
Formulating such games is at the core of modern cryptography. 
Equipped with a game that captures some security property, we can formally prove
  whether a scheme achieves that property.
%\end{figure}
%\floatstyle{plain}
%\restylefloat{figure}

The remainder of this article will explore fundamental security properties for elections, namely, ballot secrecy and verifiability.
The definitions we consider are suitable for
a large class of election systems.
And we demonstrate their applicability by reviewing  
security of Helios.

\floatstyle{plain}
\restylefloat{figure}

\section*{Ballot secrecy}

Ballot secrecy could be formulated as game $\GZero$, which proceeds as follows:
the adversary $A$ picks choices $v_0$ and $v_1$; the challenger $C$ constructs a ballot for one of these choices, that is, the challenger selects a bit $\beta$ uniformly at random and constructs a ballot, denoted $b(v_\beta)$, for choice $v_\beta$; and 
the adversary must determine which choice the ballot is for, that is, the adversary must 
determine $\gamma$ such that $\gamma = \beta$.
If the adversary wins, then 
a voter's choice can be revealed, otherwise, it cannot, 
i.e., the election scheme provides ballot secrecy.
Helios 2.0 satisfies this notion of security, because choices are protected by encryption.

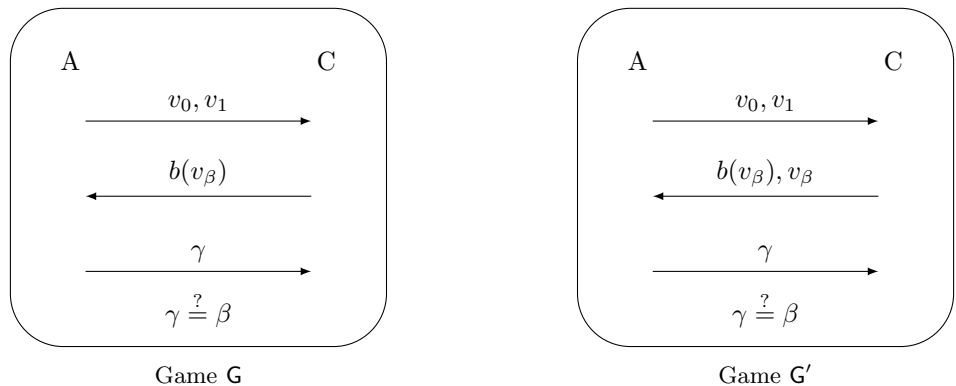
\begin{figure}[h!]
\captionsetup[subfigure]{labelformat=empty}
\begin{subfigure}{0.5\textwidth}
\centering
\begin{tikzpicture}
	\draw[rounded corners=20pt] (0,0) rectangle (5,-4.5);			
	\node at (0.8,-0.7) {A}; 
	\node at (4.2,-0.7) {C}; 
	\draw[-latex] (1,-1.5) -- node[above] {$v_0,v_1$} (4.0,-1.5);
	\draw[latex-] (1,-2.5) -- node[above] {$b(v_{\beta})$} (4.0, -2.5);
	\draw[-latex] (1,-3.5) -- node[above] {$\gamma$} (4.0,-3.5);
	\node at (2.5, -4.0) {$ \gamma \stackrel{?}{=} \beta $};
\end{tikzpicture}
\caption{Game $\GZero$}
\end{subfigure}
\begin{subfigure}{0.5\textwidth}
\centering
\begin{tikzpicture}
	\draw[rounded corners=20pt] (0,0) rectangle (5,-4.5);			
	\node at (0.8,-0.7) {A}; 
	\node at (4.2,-0.7) {C}; 
	\draw[-latex] (1,-1.5) -- node[above] {$v_0,v_1$} (4.0,-1.5);
	\draw[latex-] (1,-2.5) -- node[above] {$b(v_{\beta}), v_{\beta}$} (4.0, -2.5);
	\draw[-latex] (1,-3.5) -- node[above] {$\gamma$} (4.0,-3.5);
	\node at (2.5, -4.0) {$ \gamma \stackrel{?}{=} \beta $};
\end{tikzpicture}
\caption{Game $\GZeroPrime$}
\end{subfigure}
\end{figure}

Game $\GZero$ is too weak, because election schemes announce election outcomes
and such information can be used to reveal voters' choices.
Thus, it 
is necessary to extend the game to include some tallying capability which permits the adversary to learn the outcome.
We derive game $\GZeroPrime$ as a strengthing of game $\GZero$, whereby the challenger additionally
tallies the ballot it constructed and gives the resulting outcome to the adversary. (That is a strengthening, because there are more ways to win game $\GZeroPrime$, indeed, any adversary that wins against $\GZero$ can also win against $\GZeroPrime$, moreover, an adversary against $\GZeroPrime$ can exploit additional information -- namely, the outcome -- to win.)
However, such an outcome includes only the choice used by the challenger to construct the ballot, from which the adversary can trivially determine what choice the ballot is for. Thus, the game is 
unsatisfiable. This is inevitable, because
there are some scenarios in which outcomes reveal choices (most notably, when all voters make the same choice), as well as scenarios in which outcomes, coupled with partial knowledge on the distribution of voters' choices, allow voters' choices to be deduced. For example, suppose Alice, Bob and Mallory participate in a referendum, and 
the outcome has frequency two for \emph{yes} and one for \emph{no}. 
Mallory and Alice can deduce
Bob's choice by pooling knowledge of their own choices. Similarly, Mallory and Bob can deduce Alice's vote. Furthermore, Mallory can deduce that Alice and Bob both voted \emph{yes}, if she voted \emph{no}. 
For simplicity, our informal definition of ballot secrecy deliberately omitted
 side-conditions which 
exclude these inevitable revelations. We refine our definition as follows.

\newcommand{\secrecycondition}{the choice can be deduced
from the outcome and any partial knowledge on the distribution of
choices}

\begin{quote}
 A voter's choice is not revealed to anyone, except when \secrecycondition.
\end{quote}

\noindent
This refinement ensures the aforementioned examples are not violations of
ballot secrecy.  By comparison, if Mallory's choice is \emph{yes} and she can deduce the choice of
Alice, without knowledge of Bob's choice, then ballot secrecy is violated.

\begin{wrapfigure}{r}{.34\textwidth}
\begin{tikzpicture}
	\draw[rounded corners=20pt] (0,0) rectangle (5,-4.5);			
	\node at (0.8,-0.7) {A}; 
	\node at (4.2,-0.7) {C}; 
	\draw[-latex] (1,-1.5) -- node[above] {$v_0,v_1$} (4.0,-1.5);
	\draw[latex-] (1,-2.5) -- node[above] {$b(v_{\beta}),b(v_{1 - \beta}), \{ v_{\beta}, v_{1 - \beta} \}$} (4.0, -2.5);
	\draw[-latex] (1,-3.5) -- node[above] {$\gamma$} (4.0,-3.5);
	\node at (2.5, -4.0) {$ \gamma \stackrel{?}{=} \beta $};
\end{tikzpicture}
\caption{Game $\GOne$}
\end{wrapfigure}

We can weaken game $\GZeroPrime$,
in accordance with the refined definition of ballot secrecy, so that the adversary does not win by exploiting inevitable revelations (i.e., the adversary loses when \secrecycondition).
However, in such a  game,
the outcome includes only the choice used by the challenger to construct the ballot, from which the adversary can trivially determine what choice the ballot is for. Yet, the refined definition excludes the adversary's success in this case, since the choice is deduced from the outcome. Thus, the adversary can never win.  
We need a new approach.

We introduce game $\GOne$, which proceeds as follows.
The game is initialised by the challenger picking a bit $\beta$ uniformly at random.
The adversary picks choices $v_0$ and $v_1$, the challenger constructs a ballot for $v_\beta$, and 
gives the ballot to the adversary.
The challenger then constructs a ballot for $v_{1-\beta}$ and 
gives that ballot to the adversary too.
Thus, the challenger constructs ballots 
for $v_0$ then $v_1$, or vice-versa. The challenger tallies the two ballots and gives the resulting outcome to the adversary.
(For simplicity, we represent outcomes as multisets of choices.) The adversary must determine if $\beta = 0$ or $\beta=1$. 
This game is satisfied by Helios 2.0, because tallying ballots for 
$v_0$ and $v_1$, or ballots for $v_1$ and $v_0$, results in an 
outcome 
with frequency one for each of choices $v_0$ and $v_1$,
since the operator to combine encrypted choices is commutative.

\begin{wrapfigure}{r}{.34\textwidth}
\begin{tikzpicture}
	\draw[rounded corners=20pt] (0,0) rectangle (5,-6.5);			
	\node at (0.8,-0.7) {A}; 
	\node at (4.2,-0.7) {C}; 
	\draw[-latex] (1,-1.5) -- node[above] {$v_0,v_1$} (4.0,-1.5);
	\draw[latex-] (1,-2.5) -- node[above] {$b(v_{\beta}),b(v_{1 - \beta})$} (4.0, -2.5);
	\draw[-latex] (1,-3.5) -- node[above] {$b(v)$} (4.0,-3.5);
	\draw[latex-] (1,-4.5) -- node[above] {$\{ v_{\beta}, v_{1-\beta} , v \}$} (4.0, -4.5);
	\draw[-latex] (1,-5.5) -- node[above] {$\gamma$} (4.0,-5.5);
	\node at (2.5, -6.0) {$ \gamma \stackrel{?}{=} \beta $};
\end{tikzpicture}
\caption{Game $\GTwo$}
\end{wrapfigure}

Game $\GOne$ strengthens game $\GZero$ to include a tallying capability, 
whilst avoiding the problems associated with game $\GZeroPrime$.
However, game $\GOne$ 
is also too weak, because it does not consider that voters might be malicious or influenced by
an adversary. Thus, some attacks cannot be detected.
Indeed, Helios 2.0 is 
vulnerable to the following attack: an adversary observes a voter casting their ballot, 
casts a copy of that ballot as their own, and deduces the voter's choice from the election
outcome~\cite{Smyth12:Helios}. 
For example, in an election with voters Alice, Bob, and Mallory, if Mallory casts
a copy of Bob's ballot, then she can deduce Bob's choice as the choice
in the outcome with frequency two or greater. 

To detect the attack against Helios 2.0, we extend game $\GOne$ to (optionally) include the adversary  
picking a ballot and the challenger tallying that ballot along with the two ballots 
constructed by the challenger. We call this game $\GTwo$.
Helios 2.0 is not secure with respect to this game, because 
it detects the aforementioned attack, whereby Mallory casts a copy of Bob's ballot.
This attack can be attributed to tallying meaningfully related ballots, and omitting such ballots from tallying (i.e., ballot weeding) is intended to serve as a defence.
Indeed, the next Helios release, henceforth
\heliosspec, plans to incorporate ballot weeding to mitigate against this attack. %
As a result, \heliosspec\ satisfies the notion of ballot secrecy captured by game $\GTwo$ (cf.~\cite{Bernhard12:Helios,BCGPW15}).

In the games considered so far, the challenger always tallies the ballots it constructed
and the adversary cannot exclude those ballots from tallying. 
	This corresponds to recording all cast ballots, which introduces a trust 
	assumption on both the mechanism used to record ballots and the communication channels
	used to cast ballots.
	Attacks that arise when this trust assumption is not upheld cannot be detected. For example,
	attacks that require cast ballots to be excluded from tallying are
	not detected.
Indeed, without such an assumption, \heliosspec\ is vulnerable to an attack: 
Mallory observes Alice's ballot, 
derives a related ballot, excludes Alice's ballot from tallying, and exploits
a relationship that arises between Alice's choice and the election outcome to deduce 
Alice's choice.
This attack is similar to the attack against Helios 2.0, which involved
tallying meaningfully related ballots. The difference here is that a 
related ballot is derived and the original ballot is discarded, yet the relationship
between Alice's choice and the outcome remains, which permits the attack.
In this instance, ballot weeding is not possible, because
the original ballot is discarded. 
Nevertheless, the attack can be prevented by eliminating the possibility to construct
related ballots. 

\floatstyle{ruled}
\restylefloat{figure}
\begin{figure}\caption{Helios: Ballot construction and tallying~\cite{2018-secrecy-verifiability-elections-tutorial}}
The voter selects a choice $v$ from a list of choices $1,\dots,\ell$ and 
computes ciphertexts $\EEnc{m_1},\dots,\EEnc{m_{\ell-1}}$ such that if $v<\ell$, then 
plaintext $m_v$ is $1$ and the remaining plaintexts $m_1,\dots,m_{v-1},m_{v+1},\dots,m_{\ell-1}$
are $0$, otherwise, all plaintexts 
are $0$.
The voter also computes proofs $\sigma_1,\dots,\sigma_\ell$
so that this can be verified: proof $\sigma_1$ demonstrates $\EEnc{m_1}$
is a ciphertext on plaintext $0$ or $1$, and similarly for proofs $\sigma_2,\dots,\sigma_{\ell-1}$,
and proof $\sigma_\ell$ demonstrates that the homomorphic combination of 
ciphertexts $\EEnc{m_1},\dots,\EEnc{m_{\ell-1}}$ contains $0$ or $1$. The voter outputs
the ciphertexts and proofs as their ballot.
Hence, the ballot is for exactly one choice, and this can be verified
by checking proofs.

\newcommand{\spaceddots}{\ \ \dots \ \ }

The administrator collects ballots for which the encapsulated 
proofs all hold, forms a matrix of the encapsulated ciphertexts, 
i.e.,
\begin{align*}
	\begin{array}{r@{\hspace{3mm}}c@{\hspace{3mm}}r}
		\cellcolor{light-gray} \EEnc{m_{1,1}} & \cellcolor{light-gray} \spaceddots & \cellcolor{light-gray} \EEnc{m_{1,\ell-1}} \\
		 \multicolumn{1}{c}{\vdots}  &       & \multicolumn{1}{c}{\vdots} 	   \\
		\cellcolor{light-gray} \EEnc{m_{k,1}} & \cellcolor{light-gray} \spaceddots & \cellcolor{light-gray} \EEnc{m_{k,\ell-1}},
\intertext{homomorphically combines the ciphertexts in each column to derive
	the encrypted outcome, i.e.,} 
		\EEnc{\Sigma_{i=1}^k m_{i,1}} & \spaceddots &  \EEnc{\Sigma_{i=1}^k m_{i,\ell-1}},
\intertext{decrypts the homomorphic combinations to reveal the frequency of choices 
$1,\dots,\ell-1$, i.e.,}
		\Sigma_{i=1}^k m_{i,1}     & \spaceddots & \Sigma_{i=1}^k m_{i,\ell-1},
	\end{array}
\end{align*}
computes the frequency of choice $\ell$ by subtracting 
the frequency of any other choice from the number of collected 
ballots, i.e., $k-\Sigma_{j=1}^{\ell-1}\,\Sigma_{i=1}^k m_{i,j}$,
and announces the outcome as those frequencies,
along with a proof demonstrating correctness of decryption.
\end{figure}

\floatstyle{plain}
\restylefloat{figure}

Ballot secrecy can be formulated as a game that eliminates the undesirable assumption 
that all cast ballots are recorded 
and tallied, and considers the adversary %\sout{ and challenger} 
casting arbitrarily many ballots. The game is more complex than the games we have considered, since it must introduce non-trivial side conditions to ensure the adversary does not win by exploiting inevitable revelations, and we refer the reader to~\cite{Smyth15:ballotSecrecy} for details. 
Using this game, attacks against \heliosspec\ can be detected~\cite{Smyth15:ballotSecrecy}.
Nonetheless, 
a variant
of 
Helios~\cite{2014-election-verifiability}, henceforth Helios'16, that uses 
ballots from which meaningfully related ballots cannot be constructed (i.e., non-malleable ballots)~\cite{Smyth15:ballotSecrecy,Smyth15:NM-CPA-ciphertexts}, is not vulnerable to such attacks and is proven to satisfy this formulation of ballot secrecy.

We have introduced a variant of Helios that satisfies ballot secrecy, but ballot secrecy does not
ensure free choice when adversaries are able to communicate with voters nor when voters deviate 
from the prescribed voting procedure to follow instructions provided by adversaries. Indeed, Helios 
does not prevent a voter from showing an adversary how they computed their ballot to reveal their
choice. In the presence of such adversaries, election schemes must satisfy stronger notions of free choice, such as receipt-freeness
and coercion resistance.

\floatstyle{ruled}
\restylefloat{figure}

\begin{figure}\caption{Practicality: Defining security properties is challenging, are efforts well-placed?}
By exploring various formulations of ballot secrecy, we have seen how challenging defining a security property can be. 
Indeed, researchers toil away to get the subtle details of definitions right. Their efforts are well-placed, since 
these definitions enable 
the design of new, \emph{provably secure} schemes, as well as the discovery of vulnerabilities in existing schemes. The evolution of Helios showcases the value of such work. Indeed, vulnerabilities were discovered against Helios 2.0 and \heliosspec, and Helios'16 was developed to overcome these vulnerabilities.
But, are these efforts sufficient?

A formal security definition essentially captures a model of 
possible interactions between some scheme and an adversary, and a scheme can 
be proven to satisfy the security definition if no adversary can break security
within the context of the model. It follows that no adversary can break
security of the deployed scheme, as far as the model captures interactions
between the deployed scheme and any adversary. 
Hence, a model that underestimates adversarial capabilities may miss attacks
and a model that overestimates adversarial capabilities may report false 
attacks. For instance, game $\GTwo$ does not capture attacks by adversaries
that control the mechanism used to record ballots or the communication channel
used to cast ballots, consequently it misses attacks.
\end{figure}

\floatstyle{plain}
\restylefloat{figure}

\section*{Election verifiability}

\begin{wrapfigure}{r}{.34\textwidth}
\begin{tikzpicture}
	\draw[rounded corners=20pt] (0,0) rectangle (5,-2.5);			
	\node at (0.8,-0.7) {A}; 
	\node at (4.2,-0.7) {C}; 
	\draw[-latex] (1,-1.5) -- node[above] {$v_0,v_1$} (4.0,-1.5);
	\node at (2.5, -2.0) {$ b(v_0) \stackrel{?}{=} b(v_1) $};
\end{tikzpicture}
\caption{Game $\GThree$}
\end{wrapfigure}

To satisfy individual verifiability, each voter must be able to check that their ballot has been recorded. Thus, the recorded ballots must be publicly accessible.
In this setting, it suffices to enable voters to uniquely identify their ballot 
amongst the recorded ballots.
This property can be formulated as game $\GThree$, which proceeds as follows: 
the adversary provides any inputs necessary to construct a ballot, including a
choice $v_0$;  
the challenger constructs a ballot using those inputs; and the 
adversary and challenger repeat the process
to construct a second ballot for choice $v_1$. The adversary wins if the 
two independently constructed ballots are equal. Winning 
signifies the existence of scenarios in which 
voters cannot uniquely identify their ballot, thus voters cannot be convinced 
that their ballot is recorded.
To achieve individual verifiability, 
ballots must be constructed in a non-deterministic
manner. This can be achieved by including an encrypted choice 
in the ballot. Indeed,
Helios 2.0, \heliosspec\ and Helios'16 all achieve individual verifiability in this 
way~\cite{2014-election-verifiability}.

\begin{wrapfigure}{r}{.34\textwidth}
\begin{tikzpicture}
	\draw[rounded corners=20pt] (0,0) rectangle (5,-3.0);			
	\node at (0.8,-0.7) {A}; 
	\node at (4.2,-0.7) {C}; 
	\draw[-latex] (1.0,-1.5) -- node[above] {$\mathfrak v,b(v_1),\dots,b(v_n)$} (4.0,-1.5);
	\node at (2.5, -2.0) {$ \mathfrak v \stackrel{?}{\not=} \{v_1,\dots v_n\} \mathrel \wedge$};
	\node at (2.5, -2.5) {$ c( \mathfrak v,b(v_1),\dots,b(v_n)) \stackrel{?}{=} \top$};
\end{tikzpicture}
\caption{Game $\GFour$}
\end{wrapfigure}

To satisfy universal
verifiability, anyone must be able to check that the announced outcome 
corresponds to the choices expressed in the recorded ballots.
An election scheme's verification step is intended to perform such checks,
hence, it suffices to consider this step when formulating the security property.
Thus, universal verifiability can be formulated as game $\GFour$, which
proceeds a follows: the adversary provides inputs necessary for verification, including  outcome $\mathfrak v$ and  recorded ballots $b(v_1),\dots,b(v_n)$, and it wins if the outcome does not correspond to the choices expressed by those recorded ballots, yet verification succeeds, i.e., $c( \mathfrak v,b(v_1),\dots,b(v_n)) = {\top}$.
At first glance, Helios 2.0 appears to satisfy universal verifiability. Indeed, proofs of correct 
computation are produced in the setup, voting and tallying steps, and these 
proofs are checked during verification. Moreover, an abstract model of 
Helios 2.0 is proven to satisfy a notion of universal verifiability~\cite{Smyth10:ElectionVerifiability}. However, the 
mechanism to construct proofs in  Helios 2.0 is 
unsuitable and this gives way to vulnerabilities that enable the administrator 
to inject an arbitrary number of choices in the outcome~\cite{Bernhard12:Helios,Essex16:HeliosVerifiability}.
(The abstract model did not consider the details of the mechanism used to construct proofs, hence, 
the attack could not be detected in that model.)
\heliosspec\ is intended to mitigate against this attack. In particular, 
the mechanism to construct proofs
is modified. But, \heliosspec\ is reliant on ballot weeding, which is compatible with ballot secrecy~\cite{Smyth13:BallotIndependence}, but not with universal verifiability~\cite{2014-election-verifiability}.
Indeed, in \heliosspec, Mallory can cast a ballot for a choice related to Alice's choice in a way that only Mallory's choice is included in the outcome. Yet, verification would accept this outcome, despite Alice's choice being excluded, hence 
the announced outcome does not correspond to the choices expressed.
 This is an undesirable effect of the weeding procedure used by \heliosspec.
By comparison, Helios'16 uses non-malleable ballots, thereby avoiding the need for ballot weeding. And it is proven to satisfy universal verifiability~\cite{2014-election-verifiability}.

\floatstyle{ruled}
\restylefloat{figure}

\begin{figure}\caption{Trust administrators for secrecy, but not verifiability}
We consider election schemes in which the administrator tallies ballots. And 
definitions of ballot secrecy implicitly assume that the administrator tallies
the recorded ballots and nothing more. Thus, ballot secrecy can only be assured
when the administrator is honest. Indeed, if the administrator is dishonest, then
each recorded ballot can be tallied individually to reveal voters' choices. This
trust assumption can be weakened by distributing the administrator's role. Moreover, 
it can be eliminated in decentralised election schemes, such as~\cite{Schoenmakers99:PVSS,HRZ09,Smyth12:decentralised-voting-system}, 
for example.
Unlike ballot secrecy, individual and universal 
verifiability do not make any trust assumptions about the administrator. 
\end{figure}

\floatstyle{plain}
\restylefloat{figure}

\floatstyle{ruled}
\restylefloat{figure}
\section*{Closing remarks}

We have explored the fundamental properties that are necessary to ensure that election schemes behave as expected.
The exploration reveals how our understanding of those expectations has evolved, culminating in the emergence of formal, cryptographic definitions of properties necessary to fulfil expectations.
We have provided insights into definitions of secrecy and verifiability, allowing us to learn and appreciate the underlying intuition and technical details of these notions.

Equipped with definitions, we can build election schemes that can be proven to behave as expected.
And, as an illustrative example, we reviewed  Helios'16, which was built and proven secure in this way.
The definitions can also be used to analyse existing election schemes, and vulnerabilities have been uncovered. 
Indeed, we have described a series of vulnerabilities that were discovered during the analysis of Helios 2.0, which advanced our understanding of system behaviour and prompted the design of Helios'16. Moreover, the definitions are applicable beyond Helios. For instance, Smyth, Frink \& Clarkson~\cite{2014-election-verifiability} have shown that neither Helios-C (an extension of Helios)~\cite{Cortier14:verifiability} nor JCJ (an election scheme achieving coercion resistance)~\cite{JCJ10} satisfy the definition of universal verifiability, and propose a variant of JCJ that does.
Moreover, Smyth has shown that implementations of the mixnet variant of Helios do not satisfy universal
verifiability, and proposes a variant that satisfies both universal verifiability~\cite{Smyth18:HeliosMixnetVerifiability} and ballot secrecy~\cite{Smyth15:ballotSecrecy}.

\begin{figure}\caption{Variant of Helios with a mixnet}
The variant of Helios with a mixnet~\cite{Adida08} works as follows:
first, as per Helios, an administrator generates a public key $\pk$ and a proof of correct key construction.
Secondly, each voter selects a choice $v$, computes ciphertext $\EEnc{v}$ and 
a prove demonstrating the ciphertext is valid, and casts the 
ciphertext coupled with the proof as their ballot.  
Thirdly, the administrator collects ballots for which the encapsulated 
proof holds, inputs the encapsulated ciphertexts to a mixnet, and
decrypts the mixed ciphertexts to reveal the choices. That is,

\[
\left.
\begin{array}{c}
	\EEnc{v_1} \\
	\EEnc{v_2} \\
	\vdots	   \\
	\EEnc{v_k} 
\end{array}
\right\} 
	\hskip 2mm \emph{mixing} \hskip 2mm 
\left\{
\begin{array}{cccc}
	\EEnc{v_{\pi(1)}} 	& v_{\pi(1)}	\\
	\EEnc{v_{\pi(2)}} 	& v_{\pi(2)}	\\
	\vdots		   		& \vdots 		\\
	\EEnc{v_{\pi(k)}} 	&  v_{\pi(k)}
\end{array}
\right.
\]

\noindent
where $\pi$ is a permutation on $\{1,\dots,k\}$.
Moreover, the administrator derives the frequency of each choice
and announces the outcome as those frequencies, 
along with proofs demonstrating correctness of mixing and 
decryption.
Finally, any interested party checks that the outcome corresponds
to the decrypted choices and that all proofs verify, and voters verify that
the ballots they constructed are amongst those collected.
\end{figure}

We deliberately described cryptography and accompanying security proofs as a panacea that enables the construction of secure election 
schemes. We must now make a confession. There is more to this story: secure election schemes must be implemented in software, and we must prove that software is implemented as prescribed. In particular, secrecy requires setup, voting and tallying steps be implemented as prescribed; individual verifiability requires the voting step be implemented as prescribed; and universal verifiability requires the verification step be implemented as prescribed.
(Indeed, games $\GTwo$ and $\GThree$ both construct ballots in the prescribed manner, and
game $\GFour$ verifies ballots in the prescribed manner. Thus, any conclusions drawn from
security proofs only apply when the relevant steps are followed in the prescribed manner.)
Proving correct implementation is a lot more work. And that is still not enough. The issues go beyond technology:
``Voting is as much a perception issue as it is a technological issue. It's not enough for the result to be mathematically accurate; every citizen must also be confident that it is correct"~\cite{Schneier06:voting}.
Thus, proven secure election schemes are essential. Indeed, we have seen how vulnerabilities were discovered whilst
attempting to prove Helios secure. Yet, proven secure election schemes are not sufficient. And implementing secure election schemes remains
a significant research challenge.

Our notions of secrecy and verifiability generalise beyond elections.
For instance, definitions of ballot secrecy and election verifiability 
can be adapted to capture definitions of bid secrecy and auction verifiability,
and auction schemes satisfying bid secrecy and auction verifiability
can be derived from elections schemes satisfying analogous security 
properties~\cite{2014-Hawk-and-Aucitas-auction-schemes,Smyth15:Hawk}.
Thereby 
inaugurating the unification of auctions and elections.
Moreover, the notions generalise to other settings that require 
strong forms of integrity and privacy.

This article contributes to the science of security by sharing valuable insights into elections and by demonstrating the value that formal definitions and analysis have in building schemes guaranteed to behave as expected. In particular,  formulations of secrecy and verifiability facilitate the construction of secret, verifiable election schemes.
Indeed, we have seen how
Helios has advanced to thwart some attacks.

We hope this article aids
democracy-builders in deploying their systems
and helps educate administrators, policymakers and voters worldwide.

\paragraph{Acknowledgements.}

We are grateful to Aisha Chudasama Mahmud, Moshe Vardi and our anonymous reviewers
for feedback that helped improve this article, and to Maxime Meyer for illustrating our games using PGF/Ti\emph{k}Z.

\bibliographystyle{plain}
\bibliography{main-magazine}

\begin{thebibliography}{10}

\bibitem{Adida08}
Ben Adida.
\newblock {Helios: Web-based Open-Audit Voting}.
\newblock In {\em USENIX Security'08: 17th USENIX Security Symposium}, pages
  335--348. USENIX Association, 2008.

\bibitem{AdidaPereiraMarneffeQuisquater}
Ben Adida, {Olivier de} Marneffe, Olivier Pereira, and {Jean-Jacques}
  Quisquater.
\newblock {Electing a University President Using Open-Audit Voting: Analysis of
  Real-World Use of Helios}.
\newblock In {\em EVT/WOTE'09: Electronic Voting Technology Workshop/Workshop
  on Trustworthy Elections}. USENIX Association, 2009.

\bibitem{Alvarez10:ElectronicElections}
R.~Michael Alvarez and Thad~E. Hall.
\newblock {\em {Electronic Elections: The Perils and Promises of Digital
  Democracy}}.
\newblock {Princeton University Press}, 2010.

\bibitem{BCGPW15}
David Bernhard, V\'{e}ronique Cortier, David Galindo, Olivier Pereira, and
  Bogdan Warinschi.
\newblock {SoK: A comprehensive analysis of game-based ballot privacy
  definitions}.
\newblock In {\em S\&P'15: 36th Security and Privacy Symposium}, pages
  499--516. IEEE Computer Society, 2015.

\bibitem{Bernhard12:Helios}
David Bernhard, Olivier Pereira, and Bogdan Warinschi.
\newblock {How Not to Prove Yourself: Pitfalls of the Fiat-Shamir Heuristic and
  Applications to Helios}.
\newblock In {\em ASIACRYPT'12: 18th International Conference on the Theory and
  Application of Cryptology and Information Security}, volume 7658 of {\em
  LNCS}, pages 626--643. Springer, 2012.

\bibitem{Bertrand07:Opening:HiddenHistorySecrectBallot}
Romain Bertrand, Jean-Louis Briquet, and Peter Pels.
\newblock {Introduction: Towards a Historical Ethnography of Voting}.
\newblock In {\em {The Hidden History of the Secret Ballot}}. {Indiana
  University Press}, 2007.

\bibitem{DebraBowenCalifornia07}
Debra Bowen.
\newblock {Secretary of State Debra Bowen Moves to Strengthen Voter Confidence
  in Election Security Following Top-to-Bottom Review of Voting Systems}.
\newblock California Secretary of State, press release DB07:042, August 2007.

\bibitem{Essex16:HeliosVerifiability}
Nicholas {Chang-Fong} and Aleksander Essex.
\newblock {The Cloudier Side of Cryptographic End-to-end Verifiable Voting: A
  Security Analysis of Helios}.
\newblock In {\em ACSAC'16: 32nd Annual Conference on Computer Security
  Applications}, pages 324--335. ACM Press, 2016.

\bibitem{Cortier14:verifiability}
V{\'e}ronique Cortier, David Galindo, St\'ephane Glondu, and Malika
  Izabach{\`e}ne.
\newblock {Election Verifiability for Helios under Weaker Trust Assumptions}.
\newblock In {\em ESORICS'14: 19th European Symposium on Research in Computer
  Security}, volume 8713 of {\em LNCS}, pages 327--344. Springer, 2014.

\bibitem{Smyth12:Helios}
V\'{e}ronique Cortier and Ben Smyth.
\newblock {Attacking and fixing Helios: An analysis of ballot secrecy}.
\newblock {\em Journal of Computer Security}, 21(1):89--148, 2013.

\bibitem{GermanyCourt09}
{Use of voting computers in 2005 Bundestag election unconstitutional}, March
  2009.
\newblock Press release 19/2009
  \url{https://www.bundesverfassungsgericht.de/SharedDocs/Pressemitteilungen/EN/2009/bvg09-019.html}
  (accessed 17 Jan 2018).

\bibitem{Rop07:NetherlandsVoting}
Rop Gonggrijp and Willem-Jan Hengeveld.
\newblock {Studying the Nedap/Groenendaal ES3B Voting Computer: A Computer
  Security Perspective}.
\newblock In {\em EVT'07: Electronic Voting Technology Workshop}. USENIX
  Association, 2007.

\bibitem{Gumbel05:StealThisVote}
Andrew Gumbel.
\newblock {\em {Steal This Vote: Dirty Elections and the Rotten History of
  Democracy in America}}.
\newblock {Nation Books}, 2005.

\bibitem{HRZ09}
Fao Hao, Peter Y.~A. Ryan, and Piotr Zieli\'nski.
\newblock {Anonymous voting by two-round public discussion}.
\newblock {\em Journal of Information Security}, 4(2):62 -- 67, 2010.

\bibitem{JonesSimons12:VotingBook}
Douglas~W. Jones and Barbara Simons.
\newblock {\em Broken Ballots: Will Your Vote Count?}, volume 204 of {\em CSLI
  Lecture Notes}.
\newblock Center for the Study of Language and Information, Stanford
  University, 2012.

\bibitem{JCJ10}
Ari Juels, Dario Catalano, and Markus Jakobsson.
\newblock {Coercion-Resistant Electronic Elections}.
\newblock In David Chaum, Markus Jakobsson, Ronald~L. Rivest, and Peter~{Y. A.}
  Ryan, editors, {\em {Towards Trustworthy Elections: New Directions in
  Electronic Voting}}, volume 6000 of {\em LNCS}, pages 37--63. Springer, 2010.

\bibitem{Katz07}
Jonathan Katz and Yehuda Lindell.
\newblock {\em Introduction to Modern Cryptography}.
\newblock Chapman \& Hall/CRC, 2007.

\bibitem{Smyth12:decentralised-voting-system}
Dalia Khader, Ben Smyth, Peter Y.~A. Ryan, and Feng Hao.
\newblock {A Fair and Robust Voting System by Broadcast}.
\newblock In {\em {EVOTE'12}: {5th International Conference on Electronic
  Voting}}, volume 205 of {\em Lecture Notes in Informatics}, pages 285--299.
  Gesellschaft f{\"u}r Informatik, {2012}.

\bibitem{Smyth10:ElectionVerifiability}
Steve Kremer, Mark~D. Ryan, and Ben Smyth.
\newblock {Election verifiability in electronic voting protocols}.
\newblock In {\em ESORICS'10: 15th European Symposium on Research in Computer
  Security}, volume 6345 of {\em LNCS}, pages 389--404. Springer, 2010.

\bibitem{Lepore08}
Jill Lepore.
\newblock {Rock, Paper, Scissors: How we used to vote}.
\newblock {\em Annals of Democracy, The New Yorker}, October 2008.

\bibitem{Lijphart84}
Arend Lijphart and Bernard Grofman.
\newblock {\em {Choosing an electoral system: Issues and Alternatives}}.
\newblock Praeger, 1984.

\bibitem{2014-Hawk-and-Aucitas-auction-schemes}
Adam McCarthy, Ben Smyth, and Elizabeth~A. Quaglia.
\newblock {Hawk and Aucitas: e-auction schemes from the Helios and Civitas
  e-voting schemes}.
\newblock In {\em FC'14: 18th International Conference on Financial
  Cryptography and Data Security}, volume 8437 of {\em LNCS}. Springer, 2014.

\bibitem{Mill1830}
James Mill.
\newblock {The Ballot}.
\newblock In {\em {The Westminster Review}}, volume~13. Robert Heward, 1830.

\bibitem{OAS:HumanRights}
{American Convention on Human Rights, ``Pact of San Jose, Costa Rica''}, 1969.

\bibitem{OSCE:HumanRights}
{Document of the Copenhagen Meeting of the Conference on the Human Dimension of
  the CSCE}, 1990.

\bibitem{Smyth15:Hawk}
Elizabeth~A Quaglia and Ben Smyth.
\newblock Secret, verifiable auctions from elections.
\newblock {\em Theoretical Computer Science}, 730:44--92, 2018.

\bibitem{Saalfeld95}
Thomas Saalfeld.
\newblock {On Dogs and Whips: Recorded Votes}.
\newblock In Herbert D\"oring, editor, {\em Parliaments and Majority Rule in
  Western Europe}, chapter~16. St. Martin's Press, 1995.

\bibitem{Schneier06:voting}
Bruce Schneier.
\newblock {Voting Technology and Security}.
\newblock
  \url{https://www.schneier.com/blog/archives/2006/11/voting_technolo.html},
  2006.

\bibitem{Schoenmakers99:PVSS}
Berry Schoenmakers.
\newblock A simple publicly verifiable secret sharing scheme and its
  application to electronic voting.
\newblock In {\em CRYPTO'99: 19th International Cryptology Conference}, volume
  1666 of {\em LNCS}, pages 148--164. Springer, 1999.

\bibitem{Smyth15:ballotSecrecy}
Ben Smyth.
\newblock {Ballot secrecy: Security definition, sufficient conditions, and
  analysis of Helios}.
\newblock Cryptology ePrint Archive, Report 2015/942, 2018.

\bibitem{2018-secrecy-verifiability-elections-tutorial}
Ben Smyth.
\newblock A foundation for secret, verifiable elections.
\newblock Cryptology ePrint Archive, Report 2018/225, 2018.

\bibitem{Smyth18:HeliosMixnetVerifiability}
Ben Smyth.
\newblock {Verifiability of Helios Mixnet}.
\newblock In {\em Voting'18: 3rd Workshop on Advances in Secure Electronic
  Voting}, LNCS. Springer, 2018.

\bibitem{Smyth13:BallotIndependence}
Ben Smyth and David Bernhard.
\newblock {Ballot secrecy and ballot independence coincide}.
\newblock In {\em ESORICS'13: 18th European Symposium on Research in Computer
  Security}, volume 8134 of {\em LNCS}, pages 463--480. Springer, 2013.

\bibitem{2014-election-verifiability}
Ben Smyth, Steven Frink, and Michael~R. Clarkson.
\newblock {Election Verifiability: Cryptographic Definitions and an Analysis of
  Helios, Helios-C, and JCJ}.
\newblock Cryptology ePrint Archive, Report 2015/233 (version 20170111:122701),
  2017.

\bibitem{Smyth15:NM-CPA-ciphertexts}
Ben Smyth, Yoshikazu Hanatani, and Hirofumi Muratani.
\newblock {NM-CPA secure encryption with proofs of plaintext knowledge}.
\newblock In {\em IWSEC'15: 10th International Workshop on Security}, volume
  9241 of {\em LNCS}, pages 115--134. Springer, 2015.

\bibitem{Halderman14:EstoniaVoting}
Drew Springall, Travis Finkenauer, Zakir Durumeric, Jason Kitcat, Harri Hursti,
  Margaret MacAlpine, and J.~Alex Halderman.
\newblock {Security Analysis of the Estonian Internet Voting System}.
\newblock In {\em CCS'14: 21st ACM Conference on Computer and Communications
  Security}, pages 703--715. ACM Press, 2014.

\bibitem{ElectoralCommision07}
UK Electoral Commission.
\newblock {\em {Key issues and conclusions: May 2007 electoral pilot schemes}},
  May 2007.
\newblock
  \url{http://www.electoralcommission.org.uk/__data/assets/electoral_commission_pdf_file/0015/13218/Keyfindingsandrecommendationssummarypaper_27191-20111__E__N__S__W__.pdf}
  (accessed 7 May 2014).

\bibitem{UN:HumanRights}
{Universal Declaration of Human Rights}, 1948.

\bibitem{VoteFoundation15}
{U.S. Vote Foundation}.
\newblock {The Future of Voting: End-to-End Verifiable Internet Voting}.
\newblock Technical report, 2015.

\bibitem{Halderman10:IndiaVoting}
Scott Wolchok, Eric Wustrow, J.~Alex Halderman, Hari~K. Prasad, Arun Kankipati,
  Sai~Krishna Sakhamuri, Vasavya Yagati, and Rop Gonggrijp.
\newblock {Security Analysis of India's Electronic Voting Machines}.
\newblock In {\em CCS'10: 17th ACM Conference on Computer and Communications
  Security}, pages 1--14. ACM Press, 2010.

\bibitem{Halderman12:DCVoting}
Scott Wolchok, Eric Wustrow, Dawn Isabel, and J.~Alex Halderman.
\newblock {Attacking the Washington, {D.C.} Internet Voting System}.
\newblock In {\em FC'12: 16th International Conference on Financial
  Cryptography and Data Security}, volume 7397 of {\em LNCS}, pages 114--128.
  Springer, 2012.

\end{thebibliography}

\end{document}